\documentclass[11pt]{article}
\usepackage{geometry} 
\usepackage{amsmath}
\usepackage{graphicx}
\usepackage{amssymb}
\usepackage{theorem}
\usepackage{ifthen}
\usepackage{epstopdf}
\usepackage{enumerate}
\newcommand\thmheadfont{\upshape\bfseries}
\theoremheaderfont{\thmheadfont}
\theorembodyfont{\itshape}
\newtheorem{thm}{Theorem}
\theoremstyle{plain}
\theorembodyfont{\upshape}
\newtheorem{prop}[thm]{Proposition}
\newtheorem{cor}[thm]{Corollary}

\newtheorem{claim}[thm]{Claim}
\newtheorem{conjecture}[thm]{Conjecture}

\newenvironment{pr}[1][!*!,!]%
{\noindent{\thmheadfont Proof%
    \ifthenelse{\equal{#1}{!*!,!}}{}{%
      \normalfont\ (#1)}\\}}%
{\koniec\medskip}
{\noindent{\thmheadfont Proof%
    \ifthenelse{\equal{#1}{!*!,!}}{}{%
      \normalfont\ (#1)}\ \ \ }}%
{\koniec\medskip}
\newcommand{\koniec}{$\Box$}

 \title{Regular languages defined by first-order formulas without quantifier alternation }
 \author{Andreas Krebs, University of T\"ubingen\\Howard Straubing, Boston College}

\begin{document}

\maketitle

\begin{abstract} We give a simple new proof that regular languages defined by first-order sentences with no quantifier alteration can be defined by such sentences in which only regular atomic formulas appear. Earlier proofs of this fact relied on arguments from circuit complexity  or algebra. Our proof is much more elementary, and uses only the most basic facts about finite automata.
  \end{abstract}
  
\section{Introduction}
\subsection{Background}

This paper concerns the use of formulas of first-order logic to define properties of strings over a finite alphabet $A.$  \footnote{We use `string' and `word' interchangeably.} The description of how these formulas work will be rather informal in this introductory section, but we will be more precise later on. Variables in our formulas are interpreted as positions in a string: these should be thought of as positive integers, where the leftmost position in a string is 1. The formulas themselves are built from a base consisting of two types of atomic formulas:  The first type consists of unary predicates $a(x),$ where $a\in A,$ which we interpret to mean that the letter in position $x$ is $a.$ The other atomic formulas we call `numerical predicates'.  What makes them numerical is that they depend only on the numerical values of positions and the length of the string, and not on the letters in those positions.  So, for example, $x<y$ is a binary numerical predicate, asserting that the position represented by the variable $x$ is strictly to the left of that represented by the position $y.$ `$x$ is prime' is a unary numerical predicate.  `The length of the string is the sum of two cubes' is a 0-ary numerical predicate.  An example of a sentence ({\it i.e.,} a formula without free variables) in this logic is
$$\forall x\forall y((a(x)\wedge b(y))\rightarrow x|y).$$
This says that every position containing the letter $a$ divides every position containing the letter $b.$ So if we interpret the sentence in strings over the alphabet $A=\{a,b,c\},$ we find that $aacbcbcc$ satisfies the sentence, but $aacbbbcc$ does not, because the latter string contains an $a$ in position 2 and a $b$ in position 5.

A sentence $\phi$  in this logic therefore defines a language in $A^*$, consisting of all words $w\in A^*$ such that $w\models\phi.$  We denote by $FO[\cal N]$ the family of languages that can be defined in this way.   The symbol $\cal N$ is meant to represent the class of {\it all} numerical predicates; the input alphabet $A$ is understood.  We will often blur the distinction between sentences and the languages that they define, so we will sometimes write $w\models\phi$ as $w\in\phi,$ and use $FO[{\cal N}]$ to denote the family of sentences as well as the associated family of languages.

This formalism for defining properties of words in first-order logic is connected on the one hand to boolean circuit complexity, and on the other to the theory of finite automata and regular languages. The connection to boolean circuit complexity is this:  $FO[\cal N]$ is identical to the circuit complexity class $AC^0$ consisting of languages recognized by constant-depth polynomial-size boolean circuit families in which the $AND$- and $OR$- gates are permitted to have unbounded fan-in. \footnote{Strictly speaking, $AC^0$ is usually defined just strings over $\{0,1\}$, while our logical formalism allows us to use any finite alphabet $A$ of input symbols.  But this is not really a problem---one can, for example, adopt any fixed-length binary encoding of $A$ and thus recognize sets of strings over $A$ by circuits. The class of languages over $A$ obtained in this way does not depend on the encoding.} This equivalence was shown by Immerman~\cite{immerman}, and, independently, Gurevich and Lewis~\cite{gurevich-lewis}. Low-depth circuit complexity is one of the very few parts of computational complexity theory for which we possess unconditional superpolynomial lower bounds: the principal result in this vein is the theorem of Furst, Saxe and Sipser~\cite{fss} that PARITY--the set of bit strings in which the number of 1's is even--is not in $AC^0.$ More generally, one cannot determine in $AC^0$ whether the number of 1's in a bit string is divisible by $k,$ where $k$ is any positive integer greater than 1. 

The connection with the theory of automata is this: If the numerical predicates used in the formulas are appropriately restricted, then the languages defined by sentences are regular languages.  Indeed, much of the algebraic theory of regular languages begins with the theorem of Sch\"utzenberger~\cite{schutzenberger}, that if the only numerical predicate permitted is $<,$ then we obtain precisely the regular languages whose syntactic monoids contain no nontrivial groups.\footnote{This theorem is usually stated without reference to logic, using an equivalent formulation in which languages are defined by a kind of restricted regular expression (`star-free' languages). However the logical interpretation has been known for just as long. }We denote this class of languages by $FO[<].$

We can add some other numerical predicates and still be able to define only regular languages.  For example, we can write $x\equiv 0\pmod 3$ to say that a position is divisible by 3.  In fact, if the only numerical predicates used can themselves be defined by finite automata (in a sense that we will make precise below), then the resulting sentences define regular languages.  We can write this fact as
$$FO[{\bf Reg}]\subseteq {\bf Reg}\cap FO[\cal N].$$
Here we have used {\bf Reg}  to denote both the regular numerical predicates and the regular languages.

Barrington, {\it et. al.}~\cite{bcst}, showed that the regular languages in $AC^0$ are precisely those in $FO[{\bf Reg}],$ and thus
 $$FO[{\bf Reg}]= {\bf Reg}\cap FO[\cal N].$$
This can be viewed as a reformulation of the theorem of Furst, Saxe and Sipser cited above:  It is relatively easy to show, using either model-theoretic games or algebraic methods, that PARITY cannot be defined in $FO[{\bf Reg}],$ so the equation above implies the circuit lower bound.  On the other hand, the circuit complexity result was itself used to prove the equation, and we do not know of any other way to obtain what on the surface appears to be a result about logic and finite automata.  It is an interesting problem to exploit this logic-automaton connection obtain a different kind of proof of the equation, one that does not use the circuit lower bounds.  The present paper is intended as a contribution towards this goal.

 \subsection{The central conjecture, and the main result}
Let $A$ be a finite alphabet, fixed throughout.
$\Sigma_k[{\cal N}]$ denotes the class of languages in $A^*$ defined by   $\Sigma_k$-sentences, in which arbitrary numerical predicates are permitted. $B\Sigma_k[{\cal N}]$ denotes the boolean closure of this class of languages. $B\Sigma_k[{\bf Reg}]$ denotes the subclass in which only regular numerical predicates are permitted.  (Precise definitions will be given below.) 

\begin{conjecture}\label{conjecture.main}  Let $k>0.$

$$\Sigma_k[{\cal N}]\cap {\bf Reg}=\Sigma_k[{\bf Reg}]$$
and
$$B\Sigma_k[{\cal N}]\cap {\bf Reg}=B\Sigma_k[{\bf Reg}].$$
\end{conjecture}

  Observe that if the conjecture holds for $\Sigma_k$ sentences, then it holds as well for $\Pi_k$ sentences, by complementation.

This conjecture is explored at length in Straubing~\cite{straubing_book}.   Our main result  is that the conjecture is true when $k=1$:

\begin{thm}\label{theorem.main}
$$\Sigma_1[{\cal N}]\cap {\bf Reg}=\Sigma_1[{\bf Reg}]$$
and
$$B\Sigma_1[{\cal N}]\cap {\bf Reg}=B\Sigma_1[{\bf Reg}].$$

\end{thm}

\subsection{Related work} 

Theorem~\ref{theorem.main} is not new:
The equality for $\Sigma_1$ is quite easy and has been known for some time. Proofs appear in Péladeau~\cite{peladeau} and Straubing~\cite{straubing_book}. We give the proof below, in Section~\ref{section.pi1}, as it forms the basis for our subsequent argument.  A proof of the equality for $B\Sigma_1$ is given by Maciel, {\it et. al.}~\cite{mpt}. A very different proof, which applies not just to boolean combinations of $\Sigma_1$ sentences but to the analogue using modular quantifiers, is given in Straubing~\cite{straubing_modular}. What distinguishes our new proof from these earlier efforts is its very elementary nature. Both \cite{mpt} and \cite{straubing_modular} used rather complex arguments, relying on circuit complexity, the algebraic characterization of dot-depth 1 languages
 and Ramsey theory.  Instead, we show how to directly rewrite the defining formula for $L$ replacing the numerical predicates by regular numerical predicates, using only the most basic tools from set theory and automata theory.
 
   Charles Paperman suggested to us that it was possible to find a proof along these lines, and we were able to work out the details once we understood the differencing trick in Section~\ref{section.differencing}. The proof has already been presented in Borlido {\it et. al.,}~\cite{Borlido} as an example of an application of more general results on lattices and boolean algebras.  We thought that the very abstract setting of that paper risks obscuring the relatively simple idea behind the proof, so we are posting this more self-contained presentation.
   
  Recently, Barloy, {\it et. al.}~\cite{bcpz}, proved the central conjecture for $\Sigma_2$, with an intricate proof drawing on circuit complexity, extremal combinatorics, and algebra.
   
    \section{Notation}

Following \cite{straubing_book}, we interpret sentences in strings over a finite alphabet $A.$ More generally, we interpret formulas in which the free variables are contained in a set   ${\cal V}=\{x_1,\ldots,x_d\}$ of variables, in  ${\cal V}$-{\it structures over} $A$: These are words over the extended alphabet $A\times 2^{\cal V}$ in which each variable appears exactly once.  For example, 
$$(a,\emptyset)(b,\{x_1,x_3\})(b,\emptyset)(a,\{x_2\})(a,\emptyset)$$
is a $\{x_1,x_2,x_3\}$-structure over $A=\{a,b\}$ that satisfies the formulas $x_1=x_3, x_1<x_2,$ and $b(x_3),$ among others.

We use ${\bf x}$ to denote the $d$-tuple of variables $(x_1,\ldots,x_d).$  Thus if $\phi$ is quantifier-free. we write $\exists {\bf x}\phi$ for a $\Sigma_1$-formula.  We similarly write ${\bf a}$ to denote a $d$-tuple of letters $(a_1,\ldots,a_d)\in A^d$.  We write ${\bf a}({\bf x})$ as an abbreviation for
$$\bigwedge_{i=1}^da_i(x_i).$$

We deliberately overload notation:  If $\phi$ is a sentence and $w\in A^*,$ then we write $w\models\phi$ and $w\in\phi$ interchangeably, considering the logical formula and the language it defines as the same thing.  Similarly, we consider sets of structures and the formulas with free variables that define these sets to be the same thing.

 If ${\bf i}=(i_1,\ldots,i_d)$ is a $d$-tuple of positions in $w\in A^*,$ then we write $w({\bf i})$ to denote both the $d$-tuple of letters in these positions and the $\{x_1,\ldots,x_d\}$-structure obtained by tagging each position $i_j$ with the corresponding variable $x_j.$  For example, the $\{x_1,x_2,x_3\}$-structure exhibited above would be denoted $abbaa(2,4,2),$  but we also write $abbaa(2,4,2)=(b,a,b).$
 
A {\it numerical predicate} $N$ with free variables ${\cal V}=\{x_1,\ldots,x_d\}$ is a set of ${\cal V}$-structures with the following property:  If $w({\bf i})\in N,$ and $v$ is obtained from $w$ by changing one letter of $w,$ then $v({\bf i})\in N.$ Thus satisfaction of the numerical predicate depends only on the length of the structure and the positions associated with the variables, and not on the letters in those positions.  Thus we can write the satisfaction relation
$$w({\bf i})\models N({\bf x})$$
as
$$({\bf i};|w|)\models N({\bf x}).$$

If we view the numerical predicate as a set of words over $A\times 2^{\cal V},$ then a {\it regular} numerical predicate is simply a numerical predicate that is a regular language over this alphabet.  Regular numerical predicates are precisely those that can be expressed by first-order formulas over the base $x_i<x_j$ and $x_i\equiv 0\pmod {q}$ for $q>0.$  (See \cite{straubing_book} for a proof of this fact.)

\section{The $\Pi_1$-ceiling of a language, and the $\Pi_1$ case}\label{section.pi1}

Let $L\subseteq A^*,$ $d>0.$  We will define a quantifier-free formula $\lceil L\rceil_d.$ The idea is that $\forall {\bf x}\lceil L\rceil_d,$  where $|{\bf x}|=d,$ will serve as a kind of closest approximation to $L$ by a $\Pi_1[{\cal N}]$ sentence. We will call this the $\Pi_1$ {\it ceiling} of $L.$ (Obviously, this depends on $d.$. However, in what follows, we will usually drop the subscript $d$--it will be understood that every quantifier block, tuple of indices, tuple of letters, {\it etc.,} has size $d$.)

Given ${\bf a}\in A^d,$ we define a numerical predicate $R_{\bf a}^L({\bf x})$ by
$$({\bf i};n)\models R_{\bf a}^L({\bf x})$$
if and only if there is some $v\in L$ with $v({\bf i})={\bf a}$ and $|v|=n.$  We define $\lceil L\rceil$ to be the quantifier-free formula
$$\bigvee_{{\bf a}\in A^d}({\bf a}({\bf x})\wedge R_{\bf a}^L(\bf x)).$$

 The next Proposition gives some basic properties of the ceiling operator.
\begin{prop}\label{proposition.ceilingproperties}

\begin{enumerate}
\item $L\subseteq \forall {\bf x}\lceil L\rceil.$ 
\item If $L$ is itself defined by a $\Pi_1$ sentence, then $L=\forall {\bf x}\lceil L\rceil.$
\item If $L_1\subseteq L_2,$ then $\lceil L_1\rceil\subseteq\lceil L_2\rceil.$
\item If $L$ is a regular language, then $\lceil L\rceil$ uses only regular numerical predicates.
\end{enumerate}

\end{prop}

\begin{pr} 
\noindent 1.  Let $w\in L.$  Let ${\bf i}$ be a $d$-tuple of positions in $w.$  Then $w({\bf i})\models {\bf a}({\bf x})$ for exactly one ${\bf a}\in A^d.$  Since $w\in L,$ by definition $w({\bf i}) \models R_{\bf a}^L({\bf x}).$  Since ${\bf i}$ was arbitrary, we have $w\models\forall{\bf x}\lceil L\rceil.$

\smallskip

\noindent 2.  Now suppose $L=\forall {\bf x} \phi$ for some quantifier-free $\phi.$  We can write a formula equivalent to $\phi$ in disjunctive normal form as
$$\bigvee_{{\bf a}\in A^d}({\bf a}({\bf x})\wedge N_{\bf a}({\bf x})),$$
where each $N_{\bf a}$ is a numerical predicate.  We claim $L=\forall{\bf x}\lceil L\rceil.$  We already have inclusion from left to right.  So now suppose $w\in \forall{\bf x}\lceil L\rceil,$ and suppose, contrary to what we want to prove, that $w\notin L.$  Thus there is some $d$-tuple ${\bf i}$ of positions in $w$ such that $w({\bf i})\not\models\phi.$  Let ${\bf a}=w({\bf i}).$  Then $w({\bf i})\not\models N_{\bf a}({\bf x}).$  However, by assumption, $w({\bf i})\models R_{\bf a}^L({\bf x}).$  Thus there is some $v\in L$ with $v({\bf i})={\bf a}$ and $|v|=|w|.$ Since $v\in L$ we must then have $v({\bf i})\models N_{\bf}({\bf x}).$  But since $N_{\bf a}$ is a numerical predicate, we would then also have $w({\bf i})\models N_{\bf a}(\bf x),$ a contradiction.

\smallskip

\noindent 3. Clearly if $L_1\subseteq L_2$ then $R_{\bf a}^{L_1}({\bf x})\subseteq R_{\bf a}^{L_2}({\bf x}),$ from which the result follows.

\smallskip

\noindent 4. We have to show that if $L$ is a regular language and ${\bf a}\in A^d,$ then there is a finite automaton ${\cal A}$ over the alphabet $A\times 2^{\{x_1,\ldots,x_d\}}$ such that the set of $\{x_1,\ldots,x_d\}$-structures accepted by ${\cal A}$ is exactly $R_{\bf a}^L({\bf x}).$ To construct ${\cal A},$ begin with an automaton that recognizes $L.$  Then replace every edge
$$q\stackrel{a}{\longrightarrow}q',$$
where $a\in A,$
by all the edges
$$q\stackrel{(b,S)}{\longrightarrow} q',$$
where $b\in A,$ and 
$$S\subseteq\{x_j:a_j=a\}.$$ 
That is, we replace a single edge labeled $a$ in the original automaton by multiple edges: These can be labeled by {\it any} letter of $A.$ The variables contained in the second component $S$ must correspond to components of the tuple {\bf a} that contain the letter $a.$  The new automaton ${\cal A}$ is non-deterministic.

Keep the same initial and accepting states as in the original automaton that recognizes $L.$  The set of structures accepted by this new automaton ${\cal A}$ is $R_{\bf a}^L({\bf x}).$ The reason is that each accepting path in ${\cal A}$ labeled by a structure $v({\bf i})$ traverses the same states as an accepting path in the original automaton labeled by a word $w$ with $w({\bf i})={\bf a}.$ Our new automaton might accept words over $A\times 2^{\{x_1,\ldots,x_d\}}$ that are not structures, because they do not contain each variable exactly once.  However, the set of legal structures is itself a regular language, so we get regularity of $R_{\bf a}^L({\bf x})$ by intersection. Easily, the set of structures ${\bf a}({\bf x})$ is also regular, and it follows that $\lceil L\rceil$ is regular.
\end{pr}

\begin{cor}\label{corollary.pi1case}
$$\Pi_1[{\cal N}]\cap {\bf Reg}=\Pi_1[{\bf Reg}].$$
$$\Sigma_1[{\cal N}]\cap {\bf Reg}=\Sigma_1[{\bf Reg}].$$
\end{cor}

\begin{pr}
If a regular language $L$ is defined by a $\Pi_1$ sentence, then property 2 of Proposition~\ref{proposition.ceilingproperties} implies that $L=\forall {\bf x}\lceil L\rceil,$ and property 4 implies that this uses only regular numerical predicates.  If a regular language $L$ is defined by a $\Sigma_1$ sentence $\psi,$ then its complement is defined by $\neg\psi,$ which is equivalent to a $\Pi_1$ sentence, and the result follows from the property for $\Pi_1.$ 

\end{pr}
\section{Iterated differences of monotone formulas}\label{section.differencing}

In order to extend Corollary~\ref{corollary.pi1case} to arbitrary boolean combinations of $\Pi_1$ sentences, we will employ a kind of normal form for propositional logic, expressing every propositional formula as an iterated difference of monotone formulas.  Since the existence of this normal form is not as obvious or as well-known as other representations of propositional formulas, we will give a complete proof.  (The use of such chains of differences goes back to Hausdorff~\cite{hausdorff}--see the discussion in Borlido, {\it et. al.}~\cite{Borlido}.)

We consider propositional formulas over a finite set of variables $\{p,q,r,\ldots\}.$  A formula is {\it monotone} if in every satisfying assignment, changing the value of one variable from False to True yields another satisfying assignment.  A formula is monotone if and only if it is equivalent to a formula in which the negation symbol does not appear (this includes the formulas T  and F).   {\it Normal-form} formulas are defined recursively as follows:
\begin{enumerate}
\item Every monotone formula is in normal form.
\item If $\phi$ is monotone and $\psi$ is in normal form, then $\phi\wedge\neg\psi$ is in normal form.
\end{enumerate}
We will also write $\phi\wedge\neg\psi$ as $\phi-\psi,$ so a formula in normal form looks like
$$\phi_1-(\phi_2-(\phi_3\ldots )\cdots)).$$

\begin{prop}\label{proposition.normalform}
Every propositional formula is equivalent to a formula in normal form.
\end{prop}

\begin{pr}

We will identify propositional formulas with $k$ variables by sets of bit strings of length $k$ representing the satisfying assignments for the formula.  So, for example,
$$(p\wedge \neg q)\vee(r\wedge \neg s)$$
is identified with
$$\{1000,1001,1010,1011,0010,0110,1110\}.$$
In particular, we treat a formula as identical to its set of satisfying assignments, and so do not distinguish between equivalent formulas.

There is the obvious partial order on bit strings: $a\leq b$ if $b$ is obtained from $a$ by switching a (possibly empty) set of  0's to 1's.  Easily, monotone propositional formulas are those corresponding to sets of bit strings that are upward closed for this partial order.  If $\phi$ is a propositional formula, then $\uparrow\phi$ represents the upward closure of $\phi,$ which is then a monotone formula.  Considered as a set of bit strings, $\uparrow\phi$ is the intersection of all the monotone formulas that contain $\phi.$

We claim that the following algorithm computes the normal form:  Given a  set $X$ of bit strings, we write
$$X = \uparrow X-(\uparrow X-X).$$
If $\uparrow X = X,$ that is, if $X$ is already monotone, then $X$ is already in normal form, and we can leave off the second term.  In particular, if $X=\emptyset,$ then $X$ is monotone, so there is nothing to do. Otherwise, we apply the algorithm recursively to $X'=\uparrow X-X,$ until we reach a monotone result.
We need to prove that the algorithm terminates, which follows from the claim:

\begin{claim}
If $X\neq\emptyset,$ then 
$$\uparrow(\uparrow X-X)\subsetneq \uparrow X.$$
\end{claim}
This ensures that the sequence of  monotone formulas generated by the algorithm strictly decreases in cardinality, so we will eventually get to a place where $\uparrow X-X$ is empty.

To prove the claim,  observe that $\uparrow$ is inclusion-preserving: that is, for any sets $U,V$  of bit strings, if $U\subseteq V,$ then $\uparrow U\subseteq\uparrow V.$  Thus, in particular
$$\uparrow(\uparrow X-X)\subseteq \uparrow\uparrow X=\uparrow X.$$
To see that the inclusion is strict, we use the hypothesis that $X$ is nonempty and pick a minimal element $x$ of $X.$  Then $\uparrow X-\{x\}$ is monotone.  Combining this with  the order-preserving property, we get
$$\uparrow(\uparrow X-X)\subseteq\uparrow(\uparrow X-\{x\})=\uparrow X-\{x\}\subsetneq\uparrow X,$$
proving the claim.
\end{pr}

\noindent {\bf Example.} Let $k=3$ and let $\phi$ have satisfying assignments $101,010,111.$ We can write this formula in disjunctive normal form as 
$$(p\wedge \neg q\wedge r)\vee(\neg p\wedge q\wedge\neg r)\vee(p\wedge q\wedge r).$$
We have
$$\uparrow\phi=\{(0,1,0),(1,1,0),(0,1,1),(1,0,1),(1,1,1)\}=q\vee(p\wedge r).$$
$$\phi'=\uparrow \phi - \phi = \{(1,1,0),(0,1,1)\}.$$
$$\uparrow\phi' = \{(1,1,0),(0,1,1),(1,1,1)\}=q\wedge(p\vee r).$$
$$\uparrow\phi'-\phi'=\{(1,1,1)\}=p\wedge q\wedge r.$$
This last formula is itself monotone, so the algorithm terminates here.  Thus $\phi$ is equivalent to the normal form formula
$$(q\wedge(p\vee r))-((q\vee(p\wedge r))-(p\wedge q\wedge r)).$$

\section{Proof of the main theorem}\label{section.mainproof}

A boolean combination of $\Sigma_1$ sentences is equivalent to a propositional formula in which the atoms are replaced by sentences  of the form $\forall {\bf x}\phi,$ where $\phi$ is quantifier-free.  Now observe that a {\it monotone} formula in these atoms can be replaced by a single atom, because the class of $\Pi_1$ formulas is closed under both conjunction and disjunction. Note that for disjunction we need more variables---that is,
$$\forall {\bf x}\phi(\bf x)\vee\forall{\bf x}\psi(\bf x)\equiv\forall{\bf x}{\bf x}'(\phi({\bf x})\vee\psi({\bf x}')),$$
where ${\bf x}'$ is a new copy of the variables.  With conjunction you don't need to increase the number of variables.

Thus by our normal form theorem, every sentence $B\Sigma_1[\cal N]$ is equivalent to an iterated difference
$$\forall{\bf x}\phi_1-(\forall{\bf x}\phi_2-(\forall{\bf x} \phi_3-\cdots\forall{\bf x}\phi_k)) \cdots )).$$
We can insure that all of the blocks of variables ${\bf x}$ in this formula have the same length $d$ by adding additional dummy variables.

Let $L\subseteq A^*.$  We define  {\it $L$-derived languages} and {\it $L$-derived quantifier-free formulas} recursively as follows:
\begin{enumerate}
\item $L$ is an $L$-derived language.
\item Any boolean combination  of $L$-derived languages is an $L$-derived language.
\item If $\theta$ is an $L$-derived quantifier-free formula with free variables ${\bf x},$ then $\forall {\bf x}\theta$ is an $L$-derived language.
\item If $K$ is an $L$-derived language and $d>0$, then $\lceil K\rceil_d$ is an $L$-derived quantifier-free formula.
\end{enumerate}

Now suppose $L$ is a language in $B\Sigma_1[\cal N]$, so that
$$L=\forall{\bf x}\phi_1-(\forall{\bf x}\phi_2-(\forall{\bf x} \phi_3-\cdots\forall{\bf x}\phi_k)) \cdots )),$$
as above.  We claim that $L$ is defined by a sentence

$$\forall{\bf x}\psi_1-(\forall{\bf x}\psi_2-(\forall{\bf x} \psi_3-\cdots\forall{\bf x}\psi_k)) \cdots )),$$
where each $\psi_i$ is an $L$-derived quantifier-free formula.  This gives the result  about $B\Sigma_1[{\cal N}]\cap {\bf Reg}$ as a special case, since we showed earlier that when $L$ is regular, every $L$-derived quantifier-free formula uses only regular numerical predicates.

 We proceed by induction on $k,$ and note that the case $k=1$ was done in Section~\ref{section.pi1}.  Now suppose $k>1.$ Let
$$L'=\forall {\bf x}\lceil L\rceil.$$
Further, let
$$K_1=\forall {\bf x}\phi_1, K_2=\forall{\bf x}\phi_2-(\forall{\bf x} \phi_3-\cdots\forall{\bf x}\phi_k)) \cdots )$$
so that $L=K_1-K_2.$  Since $L\subseteq K_1,$ monotonicity shows that $L'\subseteq \forall {\bf x}\lceil K_1\rceil,$ and since $K_1$ is defined by a $\Pi_1$-sentence, this last is equal to $K_1$ (by Property 2 of the ceiling operator).  So $L'\subseteq K_1.$  We thus have
\begin{eqnarray*}
L'-L &=& L'\cap\overline{L}\\
&=& L'\cap\overline{K_1\cap\overline{K_2}}\\
&=& L'\cap(\overline{K_1}\cup K_2)\\
&=&(L'-K_1)\cup (L'\cap K_2)\\
&=& L'\cap K_2,
\end{eqnarray*}
the last line following  from our observation that $L'\subseteq K_1.$

  $L'$ is defined by a $\Pi_1$-sentence, and  $K_2$  is defined by an iterated difference with  $k-1$ $\Pi_1$ sentences as terms.  It follows that their intersection $L'\cap K_2$ is defined by an iterated difference with $k-1$ terms.  This is because of the identity
  $$p\wedge (q-r)\equiv (p\wedge q)-r,$$
  so that if $p$ and $q$ are both $\Pi_1$ sentences, then,  $p\wedge q$ is as well.    Thus $L'-L$ is defined by an iterated difference with $k-1$ terms.  By the inductive hypothesis, $L'-L$ is defined by such a difference  $\theta$ in which every quantifier-free formula is $(L'-L)$-derived, hence $L$-derived.  Thus, since $L\subseteq L',$
  $$L=L'-(L'-L)=\forall {\bf x}\lceil L\rceil -\theta,$$
  which is an iterated difference with $k$ terms, each of which uses  only $L$-derived quantifier-free formulas.

\section{Further work}\label{section.furtherwork}

We would like to extend this argument to prove the conjecture for values of $k$ greater than 1.  One obstacle to doing this is figuring out exactly how to generalize the ceiling operator.  Here is one idea that one might try: Any $\Pi_{k+1}$ sentence, where $k\geq 1,$ has the form 
$$\forall {\bf x}\phi,$$
where $\phi$ is a boolean combination of $\Sigma_k$ formulas whose free variables are contained in ${\cal V}=\{x_1,\ldots, x_d\}$.  While $\Sigma_k$ formulas in general can have arbitrarily large quantifier depth and use arbitrarily many numerical predicates, a fixed formula contains only finitely many numerical predicates, and the $\Sigma_k$ formulas have bounded quantifier depth.  Thus we will restrict our attention to formulas $\phi$ that are boolean combinations of $\Sigma_k$ formulas using numerical predicates from a fixed finite set, and whose quantifier depth is bounded above by some constant $t.$ We will call these {\it conforming} formulas: For any choice of parameters $d,t$ and a finite set of numerical predicates, there are only finitely many inequivalent conforming formulas. 
Now suppose $L\subseteq A^*,$ and consider the set of ${\cal V}$-structures
$$\hat L=\bigvee_{u\in L\\ {\bf j}\in |u|^d} F(u,{\bf j}),$$
where $F(u,{\bf j})$ is the conjunction of all the conforming formulas satisfied by $u({\bf j}).$ Since this conjunction is finite, $F(u,{\bf j})$ is itself a conforming formula. And while the disjunction is over infinitely many structures $u({\bf j}),$ there are only finitely many distinct disjuncts, so $\hat L$ itself is a conforming formula.  It is easy to show that $\hat L$ has the first three properties of our ceiling operator $\lceil L\rceil,$ enumerated in Proposition~\ref{proposition.ceilingproperties}.  However, we see no way to show that it has the crucial fourth property of preserving regularity. If it did, it would be possible to extend our argument by induction to higher levels, and prove Conjecture~\ref{conjecture.main}.

Here is a problem that may be simpler to solve:  Suppose that the first part of the conjecture

$$\Sigma_k[{\cal N}]\cap {\bf Reg}=\Sigma_k[{\bf Reg}]$$
holds for some $k.$ Can we then use the methods of this paper to extend this to the boolean closure $B\Sigma_k$?  As we mentioned earlier, there is a new proof of the conjecture for $\Sigma_2,$ so such a stratagem would settle the conjecture for all sentences with a single quantifier alternation.

One of the goals of this research is to find a different kind of proof of the theorem of Furst, Saxe and Sipser discussed in the introduction, by recasting it as a problem about automata and logic and attacking this translation of the problem directly.  This is not a terribly ambitious goal, since, after all, the theorem has already been proved in a number of different ways, the proofs are not exceptionally difficult, and actually give stronger lower bounds than would be obtained by what we propose here.  However, Barrington {\it et. al.}~\cite{bcst} also discuss the analogous identity when one allows both ordinary and modular quantifiers:
$$FOMOD[{\bf Reg}]\subseteq {\bf Reg}\cap FOMOD[\cal N],$$
and show that this is equivalent to the long-open conjecture in circuit complexity that the class $ACC^0$ is strictly smaller than $NC^1.$ Thus an approach that works directly on these reformulated versions of the circuit problems would be of considerable interest.

\bibliographystyle{plain}
\bibliography{bsigma_1v4}

 \end{document}